\documentclass[acmtoms]{acmtrans2m}

\usepackage{stmaryrd}
\usepackage{amsfonts}
\usepackage{amssymb}
\usepackage{amsmath}
\usepackage{verbatim}
\usepackage{graphicx}
\usepackage{here}

\newdef{definition}[theorem]{Definition}
\newdef{remark}[theorem]{Remark}

{\\*[1ex] \footnotesize\minipage{0.9\textwidth}\verbatim}
{\endverbatim\endminipage \\*[1ex]}

\newenvironment{code}[1]%
{\center\tabular{c}\hline\\ \footnotesize\minipage{#1\textwidth}\verbatim}%
{\endverbatim\endminipage\\ \\ \hline\endtabular\endcenter}

\newenvironment{pythoncode}[1]%
{\center\minipage{#1\textwidth}\footnotesize\hfill\it Python code \rm\vspace{0.1cm}\hrule\footnotesize\verbatim}%
{\endverbatim\hrule\normalsize\endminipage\newline\endcenter}

\newenvironment{cppcode}[1]%
{\center\minipage{#1\textwidth}\footnotesize\hfill\it C++ code \rm\vspace{0.1cm}\hrule\footnotesize\verbatim}%
{\endverbatim\hrule\normalsize\endminipage\newline\endcenter}

\newenvironment{ffccode}[1]%
{\center\minipage{#1\textwidth}\footnotesize\hfill\it Form compiler code \rm\vspace{0.1cm}\hrule\footnotesize\verbatim}%
{\endverbatim\hrule\normalsize\endminipage\newline\endcenter}

\newcommand{\vect}[1]{\boldsymbol{#1}}
\newcommand{\brac}[1]{\left( {#1} \right)}
\newcommand{\bracc}[1]{\left\{ {#1} \right\}}
\newcommand{\jump}[1]{\llbracket {#1} \rrbracket}
\newcommand{\avg}[1]{\langle {#1} \rangle}
\newcommand{\dx}{\,\mathrm{d}x}

\newcommand{\ds}{\,\mathrm{d}s}

\newcommand{\R}{\mathbb{R}}
\newcommand{\emp}[1]{\texttt{#1}}
\newcommand{\dolfin}{DOLFIN}
\newcommand{\fenics}{FEniCS}

\newcommand{\codeonly}[1]{%
\begin{center}
  \begin{tabular}{|c|}
    \hline
    Code \\
    \hline
    \hspace{10cm} \\
    \texttt{#1} \\
    \\
    \hline
  \end{tabular}
\end{center}}

\newcommand{\mathonly}[1]{%
\begin{center}
  \begin{tabular}{|c|}
    \hline
    Mathematical notation \\
    \hline
    \hspace{10cm} \\
    \texttt{#1} \\
    \\
    \hline
  \end{tabular}
\end{center}}

\DeclareMathOperator{\Div}{div}

\DeclareMathOperator{\Grad}{grad}

\title{\dolfin{}: Automated Finite Element Computing}

\author{ANDERS LOGG \\
        Center for Biomedical Computing, Simula Research Laboratory \\
        Department of Informatics, University of Oslo \\
        and \\
        GARTH N. WELLS \\
        Department of Engineering, University of Cambridge}

\begin{abstract}
  We describe here a library aimed at automating the solution of
  partial differential equations using the finite element method. By
  employing novel techniques for automated code generation, the
  library combines a high level of expressiveness with efficient
  computation. Finite element variational forms may be expressed in
  near mathematical notation, from which low-level code is
  automatically generated, compiled and seamlessly integrated with
  efficient implementations of computational meshes and
  high-performance linear algebra. Easy-to-use object-oriented
  interfaces to the library are provided in the form of a C++ library
  and a Python module.  This paper discusses the mathematical
  abstractions and methods used in the design of the library and its
  implementation. A number of examples are presented to demonstrate
  the use of the library in application code.
\end{abstract}

{}

\begin{document}

\begin{bottomstuff}
  A.~Logg,
  Center for Biomedical Computing,
  Simula Research Laboratory,
  P.O.~Box~134, 1325 Lysaker, Norway.
  Email: \url{logg@simula.no}.
  \newline
  G.N.~Wells, Department of Engineering, University of Cambridge,
  Trumpington Street, Cambridge CB2 1PZ, United Kingdom.
  Email: \url{gnw20@cam.ac.uk}.
\end{bottomstuff}

\maketitle

\section{Introduction}
\label{sec:introduction}

Partial differential equations underpin many branches of science and
their solution using computers is commonplace. Over time, the
complexity and diversity of scientifically and industrially relevant
differential equations has increased, which has placed new demands on
the software used to solve them. Many specialized libraries have
proved successful for a particular problem, but have lacked the
flexibility to adapt to evolving demands.

Software for the solution of partial differential equations is
typically developed with a strong focus on performance, and it is a
common conception that high performance may only be obtained by
specialization. However, recent developments in finite element code
generation have shown that this is only true in
part~\cite{kirby:2005,kirby:2006b,kirby:2006,logg:article:11}. Specialized
code is still needed to achieve high performance, but the specialized
code may be generated, thus relieving the programmer of time-consuming
and error-prone tasks.

\begin{figure}
  \begin{center}
    \includegraphics[width=7.0cm]{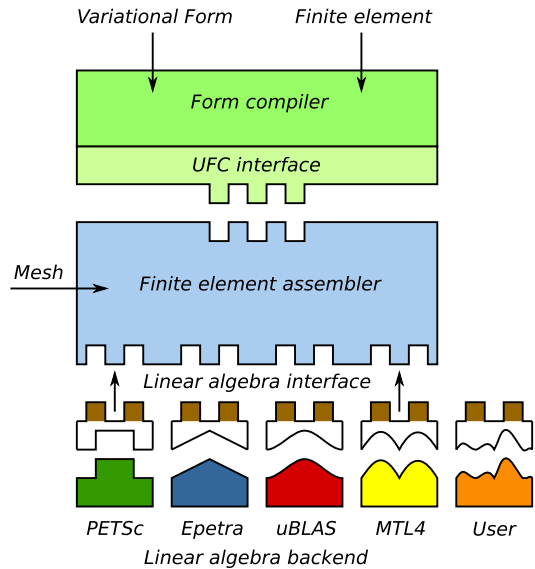}
    \caption{\dolfin{} assembles any user-defined variational form on
      any (simplex) mesh for a wide range of finite elements using any
      user-defined or built-in linear algebra backend. \dolfin{}
      relies on a form compiler for generation of the problem-specific
      code. The form compiler generates code conforming to the UFC
      (Unified Form-assembly Code) interface, either at compile-time
      or run-time, and the generated code is called during assembly by
      \dolfin{}. A small interface layer is required for each linear
      algebra backend and is implemented as part of \dolfin{} for
      PETSc, Trilinos/Epetra, uBLAS and MTL4.}
    \label{fig:variation}
  \end{center}
\end{figure}

We present in this paper the library \dolfin{} which is aimed at the
automated solution of partial differential equations using the finite
element method.  As will be elaborated, \dolfin{} relies on a
\emph{form compiler} to generate the innermost loops of the finite
element algorithm. This allows \dolfin{} to implement a general and
efficient assembly algorithm. \dolfin{} may assemble arbitrary rank
tensors (scalars, vectors, matrices and higher-rank tensors) on
simplex meshes in one, two and three space dimensions for a wide
range of user-defined variational forms and for a wide range of
finite elements. Furthermore, tensors may
be assembled into any user-defined data structure, or any of the data
structures implemented by one of the built-in linear algebra backends.
For any combination of computational mesh, variational form, finite
element and linear algebra backend, the assembly is performed by the
same code, as illustrated schematically in Figure~\ref{fig:variation},
and code generation allows the assembly code to be efficient and
compact.

\dolfin{} functions as the main programming interface and problem
solving environment of the \fenics{} Project~\cite{fenics:www}, a
collaborative effort towards the development of innovative concepts
and tools for the automation of computational mathematical modeling,
with an emphasis on partial differential equations.
See~\citeN{Logg2007a} for a overview.
All \fenics{}
components are released under the GNU General Public License or the
GNU Lesser General Public License, and are made freely available
at~\url{http://www.fenics.org}.

Initially, \dolfin{} was a monolithic, stand-alone C++ library
including implementations of linear algebra, computational meshes,
finite element basis functions, variational forms and finite element
assembly. Since then, it has undergone a number of design
iterations and some functionality has now been ``outsourced'' to other
\fenics{} components and third-party software. The design encompasses
coexistence with other libraries, and permits a user to select
particular components (classes) rather than to commit to a rigid
framework or an entire package. The design also allows \dolfin{} to
provide a complex and feature-rich system from a relatively small
amount of code, which is made possible through
automation and design sophistication.
For linear algebra functionality, third-party libraries are exploited,
with a common programming interface to these backends implemented as
part of \dolfin{}. Finite element basis functions are evaluated by
FIAT~\cite{www:FIAT,Kir04} and variational forms are handled by
the Unified Form Language (UFL) library~\cite{www:UFL,Alnaes2009}
and the
\fenics{} Form Compiler (FFC)~\cite{www:ffc,kirby:2006}.
Alternatively,
\dolfin{} may use SyFi/SFC~\cite{www:SyFi,alnaes:2009} for
these tasks, or any
other form compiler that conforms to the Unified Form-assembly Code
(UFC) interface~\cite{www:UFC} for finite element code.  Just-in-time
compilation is handled by Instant~\cite{www:Instant}.
FIAT, FFC, SyFi/SFC, UFC and Instant
are all components of the \fenics{} Project.
Data structures and algorithms for computational meshes remain
implemented as part of \dolfin{}, as is the general assembly
algorithm.

Traditional object-oriented finite element libraries, including
deal.II~\cite{bangerth:2007} and Diffpack~\cite{langtangen:book},
provide basic tools such as computational meshes, linear algebra
interfaces and finite element basis functions.  This greatly
simplifies the implementation of finite element methods, but the user
must typically implement the assembly algorithm (or at least part of it),
which
is time-consuming and error-prone. There exist today
a number of projects that seek to create systems that, at least in part,
automate the finite element method, including
Sundance~\cite{sundance:www}, GetDP~\cite{getdp:www},
FreeFEM++~\cite{freefem:www}
and Life~\cite{life:www,prudhomme2008}.
All of these rely on some form of
preprocessing (compile-time or run-time) to allow a level of
mathematical expressiveness to be combined with efficient run-time
assembly of linear systems. \dolfin{} differs from these project in
that it relies more explicitly on code generation, which allows the
assembly algorithms to be decoupled from the implementation of
variational forms and finite elements. As a result, \dolfin{} supports
a wider range of finite elements than any of the above-mentioned
libraries since it may assemble any finite element variational form on
any finite element space supported by the form compiler and finite
element backend.

The remainder of this paper is organized as follows. We first present
a background to automated finite element computing in
Section~\ref{sec:automation}. We then present some general design
considerations in Section~\ref{sec:design} before discussing in more detail
the
design and implementation of \dolfin{} in
Section~\ref{sec:implementation}. We present in
Section~\ref{sec:applications} a number of examples to illustrate the
use of \dolfin{} in application code, which is followed by concluding
remarks in Section~\ref{sec:conclusions}.

\section{Automated Finite Element Computing}
\label{sec:automation}

\dolfin{} automates the assembly of linear and nonlinear systems
arising from the finite element discretization of partial differential
equations expressed in variational form. To illustrate this, consider
the reaction--diffusion equation
\begin{equation} \label{eq:reactiondiffusion}
  -\Delta u + u = f
\end{equation}
on the unit square $\Omega = (0,1) \times (0,1)$ with $f(x, y) =
\sin(x)\cos(y)$ and homogeneous Neumann boundary conditions. The
corresponding variational problem on $V = H^{1}(\Omega)$ reads:
\begin{equation}
  \mbox{Find } u \in V : \quad a\brac{v, u} = L\brac{v} \quad \forall v \in V,
\label{eq:varproblem}
\end{equation}
where
\begin{align}
  a(v, u) &= \int_{\Omega} \nabla v \cdot \nabla u + v u \dx,
  \label{eq:varform_poisson_bilinear}
  \\
  L(v) &= \int_{\Omega} v f \dx.
  \label{eq:varform_poisson_linear}
\end{align}
To assemble and solve a linear system $A U = b$ for the degrees of
freedom $U\in\R^N$ of a finite element approximation $u_h =
\sum_{i=1}^N U_i \phi_i \in V_h \subset V$, where the set of basis
functions $\bracc{\phi_{i}}_{i=1}^N$ spans $V_{h}$, one
may simply define the bilinear~\eqref{eq:varform_poisson_bilinear} and
linear~\eqref{eq:varform_poisson_linear} forms, and then call the two
functions \emp{assemble} and \emp{solve} in \dolfin{}. This is illustrated in
Table~\ref{code:poisson} where we list a complete program for solving
the reaction--diffusion problem~\eqref{eq:reactiondiffusion} using
piecewise linear elements.

\begin{table}
\begin{pythoncode}{0.9}
from dolfin import *

mesh = UnitSquare(32, 32)
V = FunctionSpace(mesh, "CG", 1)

v = TestFunction(V)
u = TrialFunction(V)
f = Expression("sin(x[0])*cos(x[1])")

A = assemble(dot(grad(v), grad(u))*dx + v*u*dx)
b = assemble(v*f*dx)

u_h = Function(V)
solve(A, u_h.vector(), b)

plot(u_h)
\end{pythoncode}
\caption{A complete program for solving the reaction--diffusion
  problem~\eqref{eq:reactiondiffusion} and plotting the solution. This
  and other presented code examples are written for \dolfin{} version
  0.9.5 (released in December 2009).}
\label{code:poisson}
\end{table}

The example given in Table~\ref{code:poisson} illustrates the use of
\dolfin{} for solving a particularly simple equation, but assembling and
solving linear systems remain the two key steps in the solution of
more complex problems. We return to this in
Section~\ref{sec:applications}.

\subsection{Automated code generation}

\dolfin{} may assemble a variational form of any rank\footnote{Rank
  refers here to the number of arguments to the form. Thus, a linear
  form has rank one, a bilinear form rank two, etc.} from a large
class of variational forms and it does so efficiently by automated
code generation. Following a traditional paradigm, it is difficult to
build automated systems that are at the same time general and
efficient. Through automated code generation, one may
build a system which is both general and efficient.

\dolfin{} relies on a form compiler to automatically generate
code for the innermost loop of the assembly algorithm from a
high-level mathematical description of a finite element variational
form, as discussed in~\citeN{kirby:2006} and \citeN{oelgaard:2008}.
As demonstrated in~\citeN{kirby:2006}, computer code can be generated
which outperforms the usual hand-written code for a class of problems
by using representations which can not reasonably be implemented by
hand. Furthermore, automated optimization strategies can be employed
\cite{kirby:2005,kirby:2006b,logg:article:11,oelgaard:2009} and
different representations can be used, with the most efficient
representation depending on the nature of the
differential equation \cite{oelgaard:2009}. Recently, similar results
have been demonstrated in~SyFi/SFC~\cite{alnaes:2009}.

Code generation adds an extra layer of complexity to a software
system. For this reason, it is essential to isolate the parts of a
program for which code must be generated. The remaining parts may be
implemented as reusable library components in a general purpose
language. Such library components include data structures and
algorithms for linear algebra (matrices, vectors and linear/nonlinear
solvers), computational meshes, representation of functions,
input/output and plotting. However, the assembly of a linear system
from a given finite element variational formulation must be
implemented differently for each particular formulation and for each
particular choice of finite element function space(s). In particular,
the innermost loop of the assembly algorithm varies for each
particular problem. \dolfin{} follows a strategy of re-usable
components at higher levels, but relies on a form compiler to generate
the code for the innermost loop from a user-defined high-level
description of the finite element variational form.

\begin{figure}
  \begin{center}
    \includegraphics[width=9cm]{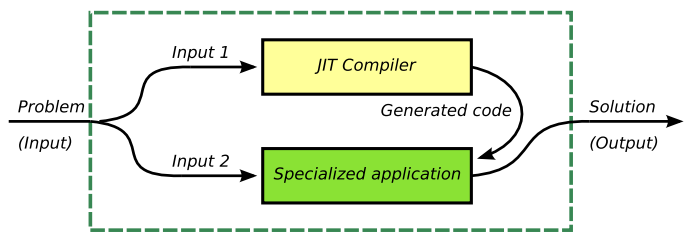}
    \caption{An automated system (\dolfin{}) using a JIT compiler
      to generate special purpose code for a subset of
      its input. For a typical finite element application, Input~1
      consists of the variational problem and the finite element(s)
      used to define it, and Input~2 consists of the mesh.}
    \label{fig:codegeneration}
  \end{center}
\end{figure}

\dolfin{} partitions the user input into two subsets: data that may
only be handled efficiently by special purpose code, and data that can
be efficiently handled by general purpose library
components. For a typical finite element application, the first set of
data may consist of a finite element variational problem and the
finite element(s) used to define it. The second set of data consists
of the mesh and possibly other parameters that define the problem. The
first set of data is given to a form compiler that generates special
purpose code. That special purpose code may then use the second set of
data as input to compute the solution. If the form compiler is
implemented as a \emph{just-in-time} (JIT) compiler, one may
seamlessly integrate the code generation into a problem solving
environment to automatically generate, compile and execute generated
code at run-time on demand. We present this process schematically in
Figure~\ref{fig:codegeneration}.

\subsection{Compilation of variational forms}

Users of \dolfin{} may use one of the two form compilers FFC or
SyFi/SFC\footnote{\dolfin{} may be used in conjunction with any form
  compiler conforming to the UFC interface.} to generate
problem-specific code. When writing a C++ application based on
\dolfin{}, the form compiler must be called explicitly from the
command-line prior to compile-time. The form compiler generates C++
code which may be included in a user program. The generated code
defines a number of classes that may be instantiated by the user and
passed to the \dolfin{} C++ library. In particular, the user may
instantiate form objects which correspond to the variational forms
given to the form compiler and which may be passed as input arguments
to the assembly function in \dolfin{}. When using \dolfin{} from
Python, \dolfin{} automatically handles the communication with the
form compiler, the compilation (and caching) of the generated code and the
instantiation of the generated form classes at run-time (JIT
compilation).

\section{Design Considerations}
\label{sec:design}

The development of \dolfin{} has been driven by two keys
factors. The first is striving for technical innovation. Examples of
this include the use of a form compiler to generate code and new data
structures for efficient representation of computational
meshes~\cite{logg:2008}. A second driving force is provided by the
needs of applications; diverse and challenging applications have
demanded and resulted in generic solutions for broad classes of
problems. Often the canonical examples have not exposed limitations in
the technology, particularly with respect to how the time required
for code generation scales with the complexity of the considered
equation \cite{oelgaard:2009}.
These have only become evident and then addressed when
attempting to solve challenging problems at the limits of current
technology. It is our experience that both these components are
necessary to drive advances and promote innovation.
We comment in this section on some generic design considerations that have been
important in the development of \dolfin{}.
\subsection{Languages and language features}
\label{sec:languages}

\dolfin{} is written primarily in C++ with interfaces provided both in
the form of a C++ class library and a Python module. The bulk of the
Python interface is generated automatically using
SWIG~\cite{swig:www,beazley:2003}, with some extensions hand-written
in Python\footnote{These extensions deal primarily with JIT
  compilation, i.e., code generation, assembly and wrapping, of
  objects before sending them through the SWIG-generated Python
  interface to the underlying C++ library.}. The Python interface
offers the performance of the underlying C++ library with the ease of
an intuitive scripting language. Performance critical operations are
developed in C++, and users can develop solvers based on \dolfin{}
using either the C++ or Python interface.

A number of C++ libraries for finite element analysis make extensive
use of templates. Templated classes afford considerable flexibility
and can be particularly useful in combining high-level abstractions
and code re-use with performance as they avoid the cost inherent in
virtual function calls in C++. However, the extensive use of templates
can obfuscate code, it increases compilation times and
compiler generated error messages are usually expansive and difficult
to interpret. We use templates in \dolfin{} where
performance demands it, and where it enables reuse of code.
However, a number of key operations in a finite element library
which require a function call involve a non-trivial number of operations
within the function, and in these cases we make use of traditional C++
polymorphism. This enhances readability and simplifies debugging
compared to template-based solutions, while not affecting run-time
performance since the extra cost of a virtual function call is
negligible compared to, for example, computing an element matrix or
inserting the entries of an element matrix into a global sparse matrix.
At the highest levels of abstraction, users are exposed to
very few templated classes and objects, which simplifies the syntax of
user-developed solvers. The limited use of templates at the user level
also simplifies the automated generation of the \dolfin{} Python
interface.

In mirroring mathematical concepts in the library design, sharing of
data between objects has proved important. For example, objects
representing functions may share a common object representing a
function space, and different function spaces may share a common
object representing a mesh. We have dealt with this issue through the
use of shared pointers, and in particular
\emp{boost::shared\_ptr} from Boost. In managing data sharing, this
solution has reduced the complexity of classes and improved the
robustness of the library.  While we make use of shared
pointers, they are generally transparent to the user and need not be
used in the high-level interface, thereby not burdening a user with
the more complicated syntax.

\subsection{Interfaces}
\label{sec:interfaces}

Many scientific libraries perform a limited number of specialized
operations which permits exposing users to a minimal, high-level
interface. \dolfin{} provides such a high-level interface for solving
partial differential equations, which in many cases allows non-trivial
problems to be solved with less than 20~lines of code (as we will
demonstrate in Section~\ref{sec:applications}). At the same time, it
is recognized that methods for solving partial differential equations
are diverse and evolving. Therefore, \dolfin{} provides interfaces of
varying complexity levels. For some problems, the minimal high-level
interface will suffice, whereas other problems may be solved using a
mixture of high- and low-level interfaces. In particular, users may often
rely on the \dolfin{} \emp{Function} class to store and hide the
degrees of freedom of a finite element function. Nevertheless, the
degrees of freedom of a function may still be manipulated directly if
desired.

The high-level interface of \dolfin{} is based on a small number of
classes representing common mathematical abstractions. These include
the classes \emp{Matrix}, \emp{Vector}, \emp{Mesh},
\emp{FunctionSpace}, \emp{Function} and \emp{VariationalProblem}. In
addition to these classes, \dolfin{} provides a small number of free
functions, including \emp{assemble}, \emp{solve} and \emp{plot}. We
discuss these classes and functions in more detail in
Section~\ref{sec:implementation}.

\dolfin{} relies on external libraries for a number of important
tasks, including the solution of linear systems. In cases where
functionality provided by external libraries must be exposed to the
user, simplified wrappers are provided. This way, \dolfin{}
preserves a consistent user interface, while allowing different
external libraries which perform similar tasks to be seamlessly
interchanged. It also permits \dolfin{} to set sensible default
options for libraries with complex interfaces that require a large
number of parameters to be set. This is most evident in the use of
libraries for linear algebra.
While the simplified wrappers
defined by \dolfin{} are usually sufficient, access is permitted to the
underlying wrapped objects so that advanced users may operate directly
on those objects when necessary.

\section{Design and Implementation}
\label{sec:implementation}

\begin{figure}
  \begin{center}
    \includegraphics[width=9cm]{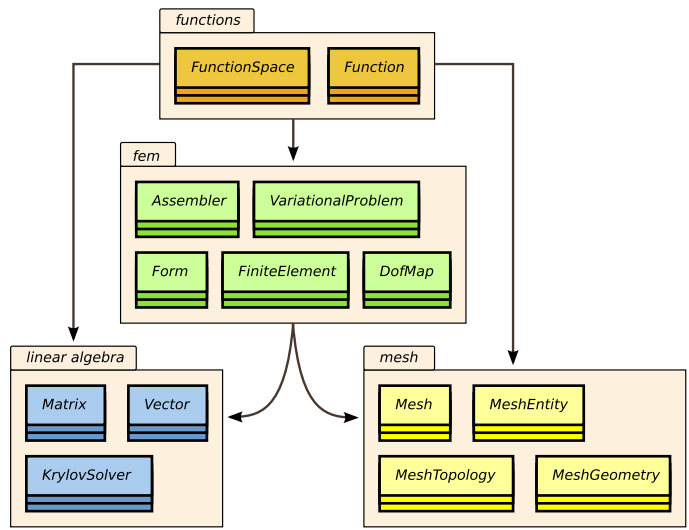}
    \caption{UML diagram of the central components and classes of \dolfin{}.}
    \label{fig:uml}
  \end{center}
\end{figure}

Like many other finite element libraries, \dolfin{} is designed as a
collection of classes partitioned into components/libraries of related
classes. However, while these classes are typically implemented as
part of the library, see, e.g.,~\citeN{bangerth:2007}, \dolfin{}
relies on automated code generation and external libraries for the
implementation of a large part of the functionality.
Figure~\ref{fig:uml} shows a UML diagram of the central components
and classes of \dolfin{}. These include the linear algebra classes,
mesh classes, finite element classes and function classes. As already
touched upon above, the linear algebra classes consist mostly of
wrapper classes for external libraries. The finite element classes
\emp{Form}, \emp{FiniteElement} and \emp{DofMap} are also wrapper
classes but for generated code, whereas the classes \emp{Assembler},
\emp{VariationalProblem} together with the mesh and function classes
are implemented as regular C++ classes (with Python wrappers) as part
of \dolfin{}.  In the
following, we address these key components of \dolfin{}, in order of
increasing abstraction. In addition to the components depicted in
Figure~\ref{fig:uml}, \dolfin{} includes a number of additional
components for input/output, logging, plotting and the solution of
ordinary differential equations.
\subsection{Linear algebra}
\label{sec:la}
\dolfin{} allows the transparent use of various specialized
linear algebra libraries. This includes the use of data structures for
sparse and dense matrices, preconditioners and iterative solvers, and
direct linear solvers. This approach allows users to leverage the
particular strengths of different libraries through a simple and
uniform interface. Currently supported linear algebra backends include
PETSc~\cite{petsc:www}, Trilinos/Epetra~\cite{trilinos:2005},
uBLAS\footnote{Krylov solvers and preconditioners for uBLAS are
  implemented as part of \dolfin{}.}~\cite{ublas:www} and
MTL4~\cite{mtl4:www}. Interfaces to the direct solvers
UMFPACK~\cite{davis:2004} (sparse LU decomposition) and
CHOLMOD~\cite{chen:2008} (sparse Cholesky decomposition) are also
provided.

The implementation of the \dolfin{} linear algebra interface is based
on C++ polymorphism. A common abstract base class
\emp{Generic\-Matrix} defines a minimal matrix interface suitable for
finite element assembly, and a subclass of \emp{Generic\-Matrix}
implements the functionality for each backend by suitably wrapping
native data structures of its respective backend. Similarly, a common
abstract base class \emp{GenericVector} defines a minimal interface
for vectors with subclasses for all backends. The two interface
classes \emp{GenericMatrix} and \emp{GenericVector} are themselves
subclasses of a common base class \emp{GenericTensor}. This enables
\dolfin{} to implement a common assembly algorithm for all matrices,
vectors and scalars (or any other rank tensor) for all linear algebra
backends.
Compared to a template-based solution, polymorphism may incur overhead
associated with the cost of resolving virtual function calls. However,
the most performance-critical function call to the linear algebra
backend during assembly is typically insertion of a local element
matrix into a global sparse matrix. This operation usually involves
a considerable amount of
computation/memory access, hence the extra cost of the virtual function
call in this case may be neglected. For cases in which the
overhead of a virtual function call is not negligible, operating directly on
the underlying object avoids this overhead.
\subsection{Meshes}

The \dolfin{} \emp{Mesh} class is based on a simple abstraction that
allows dimension-independence, both in the implementation of the
\dolfin{} mesh library and in user code. In particular, the \dolfin{}
assembly algorithm is common for all simplex meshes in one, two and
three space dimensions. We provide here an overview of the \dolfin{}
mesh implementation and refer to \citeN{logg:2008} for details.
While only simplices are currently supported, the design
paradigm extends to non-simplicial meshes.

A \dolfin{} mesh consists of a collection of \emph{mesh entities} that
define the topology of the mesh, together with a geometric mapping
embedding the mesh entities in $\R^n$. A mesh entity is a pair
$(d, i)$, where $d$ is the topological dimension of the mesh entity and $i$
is a unique index of the mesh entity. A similar approach may be found
in~\citeN{KnepleyKarpeev07A}. Mesh entities are numbered within each
topological dimension from~$0$ to $n_{d}-1$, where $n_{d}$ is the number
of mesh entities of topological dimension $d$. For convenience, mesh
entities of topological dimension~$0$ are referred to as
\emph{vertices}, entities of dimension~$1$ \emph{edges}, entities of
dimension~$2$ \emph{faces}, entities of \emph{codimension}~$1$
\emph{facets} and entities of codimension~$0$ \emph{cells}. These
concepts are summarized in Table~\ref{tab:entities}.
\begin{table}
  \begin{center}
    \begin{tabular}{|l|l|c|c|}
      \hline
      Entity & Class & Dimension & Codimension \\
      \hline
      \hline
      Mesh entity & \emp{MeshEntity} & $d$       & $D - d$ \\
      \hline
      Vertex      & \emp{Vertex}     & $0$       & $D$     \\
      Edge        & \emp{Edge}       & $1$       & $D - 1$ \\
      Face        & \emp{Face}       & $2$       & $D - 2$ \\
      \hline
      Facet       & \emp{Facet}      & $D - 1$   &  $1$    \\
      Cell        & \emp{Cell}       & $D$       &  $0$    \\
      \hline
    \end{tabular}
    \caption{\dolfin{} mesh abstractions and corresponding classes.
      Users may refer to a mesh entity either by a topological
      dimension and index or as a named mesh entity such as a vertex
      with a specific index.}
    \label{tab:entities}
  \end{center}
\end{table}

Algorithms operating on a mesh can often be expressed in terms of
\emph{iterators} \cite{Ber02,Ber06}. The mesh library provides the
general iterator \texttt{MeshEntityIterator} in addition to the
specialized mesh iterators \texttt{Vertex\-Iterator},
\texttt{EdgeIterator}, \texttt{Face\-Iterator}, \texttt{FacetIterator}
and \texttt{Cell\-Iterator}. We illustrate the use of iterators in
Table~\ref{tab:iterators}.
\begin{table}
  \begin{center}
    \begin{cppcode}{0.9}
Mesh mesh("mesh.xml");

for (CellIterator cell(mesh); !cell.end(); ++cell)
  for (VertexIterator vertex(*cell); !vertex.end(); ++vertex)
      cout << vertex->dim() << " " << vertex->index() << endl;
    \end{cppcode}
    \begin{pythoncode}{0.9}
mesh = Mesh("mesh.xml")

for cell in cells(mesh):
    for vertex in vertices(cell):
        print vertex.dim(), vertex.index()
    \end{pythoncode}
    \caption{Basic use of \dolfin{} mesh iterators for iterating
      over all vertices of all cells of a mesh in C++ (top) and
      Python (bottom).}
    \label{tab:iterators}
  \end{center}
\end{table}

The \dolfin{} mesh library also introduces the concept of a \emph{mesh
  function}. A mesh function, and its corresponding implementation
\emp{MeshFunction}, is a discrete function on the set of mesh entities
of a specific dimension. It is only defined on a set of mesh entities
which is in contrast to functions represented by the \dolfin{}
\emp{Function} class which take a value at each point in the
domain covered by the mesh.  The class \emp{MeshFunction} is templated
over the value type which allows users, for example, to create a
boolean-valued mesh function over the cells of a mesh to indicate
regions for mesh refinement, an integer-valued mesh function on
vertices to indicate a mapping from local to global vertex numbers
or a float-valued mesh function on cells to indicate material data.

The simple object-oriented interface of the \dolfin{} mesh library is
combined with efficient storage of the underlying mesh data
structures. Objects like vertices, edges and faces are never
stored. Instead, \dolfin{} stores all mesh data in plain C/C++~arrays
and provides \emph{views} of the underlying data in the form of the
class \emp{MeshEntity} and its subclasses \emp{Vertex}, \emp{Edge},
\emp{Face}, \emp{Facet} and \emp{Cell}, together with their
corresponding iterator classes. An earlier version of the \dolfin{}
mesh library used a full object-oriented model also for storage, but the
simple array-based approach has reduced storage requirements and improved the
speed of accessing mesh data by orders of magnitude~\cite{logg:2008}.
In its initial state, the \dolfin{} \emp{Mesh} class only stores
vertex coordinates, using a single array of \emp{double} values, and
cell--vertex connectivity, using a compressed row-like data structure
consisting
of two arrays of \emp{unsigned~int} values. Any other connectivity,
such as, vertex--vertex, edge--cell or cell--facet connectivity,
is automatically generated and stored when required.
Thus, if a user solves a partial differential equation using
piecewise linear elements on a tetrahedral mesh, only cell-vertex
connectivity is required and so edges and faces are not
generated. However, if quadratic elements are used, edges are
automatically generated and cubic elements will lead to a generation
of faces as well as edges.

In addition to efficient representation of mesh data, the \dolfin{}
mesh library implements a number of algorithms which operate on
meshes, including adaptive mesh refinement
(using a \citeN{rivara:1991}-type method), mesh coarsening and mesh
smoothing.
\dolfin{} does not provide
support for mesh generation, except for a number of simple shapes like
squares, boxes and spheres. The following code illustrates adaptive
mesh refinement in \dolfin{}:
\begin{cppcode}{0.9}
MeshFunction<bool> cell_markers(mesh, mesh.topology().dim());

for (CellIterator cell(mesh); !cell.end(); ++cell)
{
  if (...)
    cell_markers[*cell] = true;
  else
    cell_markers[*cell] = false;
}

mesh.refine(cell_markers);
mesh.smooth();
\end{cppcode}

\subsection{Finite elements}

\dolfin{} supports a wide range of finite elements.  At present, the
following elements are supported:
\begin{enumerate}
\item
  $H^1$-conforming finite elements:
  \begin{enumerate}
  \item
    $\mathrm{CG}_q$, arbitrary degree continuous Lagrange elements.
  \end{enumerate}
\item
  $H(\mathrm{div})$-conforming finite elements:
  \begin{enumerate}
  \item
    $\mathrm{RT}_q$, arbitrary degree Raviart--Thomas elements~\cite{RavTho77b};
  \item
    $\mathrm{BDM}_q$, arbitrary degree Brezzi--Douglas--Marini elements~\cite{BreDou85}; and
  \item
    $\mathrm{BDFM}_q$, arbitrary degree
    Brezzi--Douglas--Fortin--Marini elements~\cite{BreDou87}.
  \end{enumerate}
\item
  $H(\mathrm{curl})$-conforming finite elements:
  \begin{enumerate}
  \item
    $\mathrm{NED}_q$, arbitrary degree N\'ed\'elec elements (first
    kind)~\cite{Ned80}.
  \end{enumerate}
\item
  $L^2$-conforming finite elements:
  \begin{enumerate}
  \item
    $\mathrm{DG}_q$, arbitrary degree discontinuous Lagrange elements; and
  \item
    $\mathrm{CR}_1$, first degree Crouzeix--Raviart\footnote{Crouzeix--Raviart elements are sometimes referred to as $C^0$-\emph{non}conforming.} elements~\cite{CroRav73}.
  \end{enumerate}
\end{enumerate}
Arbitrary combinations of the above elements may be used to define
mixed elements. Thus, one may for example define a Taylor--Hood
element by combining a vector-valued $P_2$ element with a
scalar $P_1$ element. Arbitrary nesting is supported, thus
allowing a mixed Taylor--Hood element to be used as a building block
in a coupled problem which involves more than just the
velocity and pressure fields.
In Section~\ref{sec:applications},
we demonstrate the use of mixed elements for the
Poisson equation. Presently, \dolfin{} only supports elements defined
on simplices. This is not a technical limitation in the library
design, but rather a reflection of current user demand.

\dolfin{} relies on a form compiler such as FFC for the implementation
of finite elements. FFC in turn relies on FIAT for tabulation of
finite element basis functions on a reference element. In particular,
for any given element family and degree~$q$ from the list of supported
elements, FFC generates C++ code conforming to a common interface
specification for finite elements which is part of the UFC
interface. Thus, \dolfin{} does not include a library of finite
elements, but relies on automated code generation, either prior to
compile-time or at run-time, for the implementation of finite
elements. The generated code may be used for efficient run-time
evaluation of finite element basis functions, derivatives of basis
functions and evaluation of degrees of freedom (applying the
functionals to any given function). However, these
functions are rarely accessed by users as a user is not usually
exposed to the details of a finite element beyond its declaration, and
since \dolfin{} automates the assembly of variational forms based on
code generation for evaluation of the element matrix. Detailed aspects
of automated finite element code generation can be found in
\citeN{oelgaard:2008} for discontinuous elements and in
\citeN{rognes:2008} for $H({\rm div})$ and $H({\rm curl})$ elements.
\subsection{Function spaces}

The concept of a function space plays a central role in the
mathematical formulation of finite element methods for partial
differential equations. \dolfin{} mirrors this concept in the class
\emp{FunctionSpace}. This class defines a finite dimensional function
space in terms of a \emp{Mesh}, a \emp{FiniteElement} and a
\emp{DofMap} (degree of freedom map):
\begin{cppcode}{0.9}
class FunctionSpace
{
public:
  ...
private:
  ...
  boost::shared_ptr<const Mesh> _mesh;
  boost::shared_ptr<const FiniteElement> _element;
  boost::shared_ptr<const DofMap> _dofmap;
};
\end{cppcode}
The mesh defines the domain, the finite element defines the local
basis on each cell and the degree of freedom map defines how local
function spaces are patched together to form the global function
space.

For some
problems, finite element spaces are not appropriate and
a ``quadrature function space''
can be used. In such a ``space'', functions can be evaluated at discrete
points (quadrature points) but not elsewhere, and derivatives cannot
be computed. This concept is discussed in \citeN{oelgaard:2008}
and \citeN{oelgaard:2009}.

Incorporating the mathematical concept of function spaces in the
library design provides a powerful abstraction, especially for sharing
data in a transparent and simple fashion. In particular, several
functions may share the same function space and thus the same mesh,
finite element and degree of freedom mapping.

\subsection{Functions}

Functions on a finite element function space are implemented in
\dolfin{} in the form of the \emp{Function} class.  A \emp{Function}
is expressed as a linear combination of basis functions on a discrete
finite element or quadrature space. The expansion coefficients
(degrees of freedom) of the \emp{Function} are stored as a
\emp{(Generic)Vector}:
\begin{cppcode}{0.9}
class Function
{
public:
  ...
private:
  ...
  boost::shared_ptr<const FunctionSpace> _function_space;
  boost::shared_ptr<GenericVector> _vector;
};
\end{cppcode}

A \emp{Function} may be evaluated at arbitrary points on a finite
element mesh, used as a coefficient in a variational form, saved to
file for later visualization or plotted directly from within
\dolfin{}. The \dolfin{} \emp{Function} class is particularly powerful
for supplying and exchanging data between different models in coupled
problems, as will be demonstrated in Section~\ref{sec:applications}.

Evaluation of \emp{Function}s at arbitrary points is handled
efficiently using the GNU Triangulated Surface
Library~\cite{gts:www}. With the help of GTS, \dolfin{} locates which
cell of the \emp{Mesh} of the \emp{FunctionSpace} contains the given
point. The function value may then be computed by evaluating the
finite element basis functions at the given point (using the
\emp{FiniteElement} of the \emp{FunctionSpace}) and multiplying with
the appropriate coefficients in the \emp{Vector} (determined using the
\emp{DofMap} of the \emp{FunctionSpace}).

\subsection{Expressions}

Many times, it is appropriate to express a coefficient in a
variational problem by an expression or an algorithm for evaluating
the coefficient at a given point, rather than expressing it as a
linear combination of basis functions (as in the \emp{Function}
class).  Such coefficients may be conveniently implemented using the
\emp{Expression} class.

An \emp{Expression} is defined by a user through overloading the
\emp{Expression::eval} function. This \emph{functor} construct
provides a powerful mechanism for defining complex coefficients. In
particular, the functor construct allows a user to attach data to an
\emp{Expression}. A user may, for example, read data from a file
in the constructor of an \emp{Expression} subclass which is then later
accessed in the \emp{eval} callback function.

For the definition of functions given by simple expressions, like
$f(x) = \sin(x)$ or $g(x, y) = \sin(x) \cos(y)$, the \dolfin{} Python
interface provides simple and automated JIT compilation of
expressions. While the Python interface does allow a user to overload
the \emp{eval} function from Python\footnote{SWIG supports
  cross-language polymorphism using the \emph{director} feature.},
this may be inefficient as the call to \emp{eval} involves a callback
from C++ to a Python function and this may be called repeatedly during
assembly (once or more on each cell). However, JIT compilation avoids
this by automatically generating, compiling, wrapping and linking C++
subclasses of the \emp{Expression} class.

An \emp{Expression} may be evaluated at arbitrary points on a finite
element mesh, used as a coefficient in a variational form, projected
or interpolated into a finite element function space or plotted
directly from within \dolfin{}. Table~\ref{code:function} illustrates
use of the \dolfin{} \emp{Function} and \emp{Expression} classes in
Python.

\begin{table}
  \begin{center}
    \begin{pythoncode}{0.9}
# Create mesh
mesh = UnitSquare(32, 32)

# Define an expression
f = Expression(("sin(x[0])", "cos(x[1])"))

# Project expression to a finite element space
V = VectorFunctionSpace(mesh, "CG", 2)
g = project(f, V)

# Evaluate expression and function
print f(0.1, 0.2)
print g(0.1, 0.2)

# Plot expression and function
plot(f, mesh=mesh)
plot(g)
    \end{pythoncode}
    \caption{Defining, projecting, evaluating and plotting expressions
      and functions using the \dolfin{} Python interface.}
    \label{code:function}
  \end{center}
\end{table}

\subsection{Variational forms}

\dolfin{} allows general variational forms to be expressed in a form
language that mimics mathematical notation. For example, consider
the bilinear form of the standard Stokes variational problem. This
may be conveniently expressed in the form language as illustrated in
Table~\ref{tab:codevsmath}.
\begin{table}
  \mathonly{$a(v, u) = \int_{\Omega} \Grad v \cdot \Grad u - \Div v \,  p + q \, \Div u \dx$}
  \codeonly{a = (inner(grad(v), grad(u)) - div(v)*p + q*div(u))*dx}
  \caption{Expressing the bilinear form for the Stokes equations in \dolfin{}.}
  \label{tab:codevsmath}
\end{table}
The form language allows the expression of general multilinear forms
of arity~$\rho$ on the product space $V_h^1 \times V_h^2 \times \cdots
\times V_h^{\rho}$ of a sequence $\{V_h^j\}_{j=1}^{\rho}$ of finite
element spaces on a domain~$\Omega \subset \R^n$,
\begin{equation} \label{eq:a}
  a : V_h^1 \times V_h^2 \times \dots \times V_h^{\rho} \rightarrow \R.
\end{equation}
Such forms are fundamental building blocks in linear and nonlinear
finite element analysis. In particular, linear forms ($\rho=1$) and
bilinear forms ($\rho=2$) are central to the finite element
discretization of partial differential equations. Forms of higher
arity are also supported as they may sometimes be of interest,
see~\citeN{kirby:2006}.

\dolfin{} relies on the Unified Form Language
(UFL)~\cite{www:UFL,Alnaes2009} for the expression of variational forms.
The form language allows the expression of a wide range of finite
element variational forms in a language close to mathematical
notation. UFL also supports functional differentiation of general
nonlinear forms. Forms can involve integrals over cells, interior
facets and exterior facets. Line and surface integrals which do not
coincide with cell facets are not yet supported, although extensions
in this direction for modeling crack propagation have been made
\cite{nikbakht:2009}. For details on the form language, we refer to
the UFL user manual~\cite{www:UFL}.

A user of the \dolfin{} C++ interface will typically define a set of
forms in one or more form files and call FFC on the command-line. The
generated code may then be included in the user's C++ program. As an
illustration, consider again the bilinear form of the Stokes problem
as expressed in Table~\ref{tab:codevsmath}. This may be entered
together with the corresponding linear form \emp{L = v*f*dx} in a text
file named \emp{Stokes.ufl} which may then be compiled with FFC:
\begin{code}{0.9}
ffc -l dolfin Stokes.ufl
\end{code}
This will generate a C++ header file \emp{Stokes.h} which a user may
include in a C++ program to instantiate the pair of forms:
\begin{cppcode}{0.9}
#include <dolfin.h>
#include "Stokes.h"
...
int main()
{
  ...
  Stokes::FunctionSpace V(mesh);
  Stokes::BilinearForm a(V, V);
  Stokes::LinearForm L(V);
  ...
}
\end{cppcode}
When used from Python, form compilation is handled automatically by
\dolfin{}. If a form is encountered during the execution of a program,
the necessary C++ code is automatically generated and compiled. The
generated object code is cached so that code is generated and compiled
only when necessary. Thus, if a user solves the Stokes problem twice,
code is only generated the first time, as the JIT compiler will
recognize the Stokes form on subsequent runs.

\subsection{Finite element assembly}

Given a variational form, the \dolfin{} \emp{assemble} function
assembles the corresponding global tensor. In particular, a matrix is
assembled from a bilinear form, a vector is assembled from a linear
form, and a scalar value is assembled from a rank zero form (a
functional). While \dolfin{} does not provide data structures for
sparse tensors of rank greater than two, the abstract
\emp{GenericTensor} interface, which was
introduced in Section~\ref{sec:la},
permits users to supply data  structures for arbitrary rank tensors.

To discretize the multilinear form~(\ref{eq:a}), we may introduce a
basis $\{\phi^j_k\}_{k=1}^{N_j}$ for each function space $V^{j}_{h}$,
$j=1, 2, \ldots, \rho$, and define the global tensor
\begin{equation}
  A_i = a(\phi_{i_1}^1, \phi_{i_2}^2, \dots,
  \phi_{i_{\rho}}^{\rho}),
\end{equation}
where $i = (i_1, i_2, \dots, i_{\rho})$ is a multi-index.  If the
multilinear form is defined as an integral over $\Omega =
\cup_{K\in\mathcal{T}_h} K$, the tensor~$A$ may be computed by
assembling the contributions from all elements,
\begin{align}
  A_{i} = a(\phi_{i_1}^1, \phi_{i_2}^2, \dots,
  \phi_{i_{\rho}}^{\rho})
  = \sum_{K \in \mathcal{T}} a^K(\phi_{i_1}^1,
  \phi_{i_2}^2, \dots, \phi_{i_{\rho}}^{\rho}),
\end{align}
where $a^K$ denotes the contribution from element $K$. We further let
$\{\phi_k^{K,j}\}_{k=1}^{n_j}$ denote the local finite element basis
for $V^j_h$ on $K$ and define the \emph{element tensor}~$A^K$ (the
``element stiffness matrix'') by
\begin{equation}
  A_i^K = a^K(\phi_{i_1}^{K, 1},
  \phi_{i_2}^{K, 2}, \dots, \phi_{i_{\rho}}^{K, \rho}).
\end{equation}
The assembly of the global tensor~$A$ thus reduces to the computation
of the element tensor $A^K$ on each element $K$ and the insertion of
the entries of $A^K$ into the global tensor~$A$.
In addition to contributions from all cells, \dolfin{} also assembles
contributions from all exterior facets (facets on the boundary)
and all interior facets if required.

The key to the generality and efficiency of the \dolfin{} assembly
algorithm lies in the automated generation of code for the evaluation
of the element tensor. \dolfin{} relies on generated code both for the
evaluation of the element tensor and the mapping of degrees of
freedom. Thus, the assembly algorithm may call the generated code on
each cell of the mesh, first to compute the element tensor and then
again to compute the local-to-global mapping by which the entries of
the element tensor may be inserted into the global tensor.
The complexity inherent in non-trivial forms, such as those which involve
mixed element spaces, vector elements and discontinuous Galerkin methods,
is not
exposed in the form abstraction. \dolfin{} is unaware of how the
element matrix is represented or how forms are integrated. It simply
provides coefficient and mesh data to the generated code and assembles
the computed results. The algorithm for computing the element tensor
is instead determined by the form compiler. Various algorithms are
possible, including both quadrature and a special tensor
representation, and the most efficient algorithm can depend heavily on
the nature of the form \cite{kirby:2006,oelgaard:2009}.

To assemble a \emp{Form~a}, a user may simply call the function
\emp{assemble} which computes and returns the corresponding tensor.
Thus, a bilinear form may be assembled by
\begin{cppcode}{0.9}
  Matrix A;
  assemble(A, a);
\end{cppcode}
in C++ and
\begin{pythoncode}{0.9}
  A = assemble(a)
\end{pythoncode}
in Python. Several optional parameters may be given to specify either
assembly over specific subdomains of the mesh or reuse of tensors.

\subsection{Boundary conditions}

Natural boundary conditions are enforced weakly as part of
a variational problem and are typically of Neumann or Robin type, but
may also be of Dirichlet type as will be demonstrated
Section~\ref{sec:applications}. Essential boundary conditions are
typically of Dirichlet type and are enforced strongly at the linear
algebra level. \dolfin{} also supports the specification of periodic
boundary conditions. We describe here the definition and application
of strong Dirichlet boundary conditions.

We define a Dirichlet boundary condition in terms of a function
space~$V$, a function~$g$ and a subset of the boundary $\Gamma
\subseteq \partial\Omega$,
\begin{equation}
  u(x) = g(x) \quad \forall x \in \Gamma.
\end{equation}
The corresponding definition in the \dolfin{} Python interface reads
\begin{pythoncode}{0.9}
bc = DirichletBC(V, g, gamma)
\end{pythoncode}
where \emp{V} is a \emp{FunctionSpace}, \emp{g} is a \emp{Function} or
\emp{Expression}, and \emp{gamma} is a
\emp{SubDomain}. Alternatively, the boundary may be defined in terms
of a \emp{MeshFunction} marking a portion of the facets on the mesh
boundary.  The function space \emp{V} defines the space to which the
boundary condition will be applied. This is useful when applying a
Dirichlet boundary condition to particular components of a mixed or
vector-valued problem.

Once a boundary condition has been defined, it can be applied in one of two
ways. The simplest is to act upon the assembled global system:
\begin{pythoncode}{0.9}
bc.apply(A, b)
\end{pythoncode}
For each degree of freedom to be constrained, this call will zero the
corresponding row in the matrix, set the diagonal entry to one and put
the Dirichlet value at the corresponding position in the right-hand side
vector. An optional argument can be provided for updating the boundary
conditions inside a Newton iteration.  Alternatively, the
boundary condition may be supplied directly to the assembler which
will then apply the boundary condition by modifying the element matrices
in a manner that preserves any symmetry of the global matrix:
\begin{pythoncode}{0.9}
A, b = assemble_system(a, L, bc)
\end{pythoncode}

\subsection{Variational problems}

At the highest level of abstraction, objects may be created that
represent variational problems of the canonical
form~\eqref{eq:varproblem}. Such a variational problem may be defined
and solved by
\begin{pythoncode}{0.9}
problem = VariationalProblem(a, L)
u = problem.solve()
\end{pythoncode}
A constraint on the trial space in the form of one or more Dirichlet
conditions may be supplied as additional arguments. Other parameters
include the specification of the linear solver and preconditioner
(when appropriate) and whether or not the variational problem is
linear. In the case of a nonlinear variational problem where
one seeks to satisfy
\begin{equation}
  F(v) = 0 \quad \forall \ v \in V,
\end{equation}
the bilinear form~$a$ is interpreted as the Gateaux derivative of a
nonlinear form $L=F$.

\subsection{File I/O and visualization}

\dolfin{} provides input/output for objects of all its central
container classes, including \emp{Vector}, \emp{Matrix}, \emp{Mesh} and
\emp{MeshFunction}. Objects are stored to file in XML format. For
example, a \emp{Mesh} may be loaded from and stored back to file by
\begin{cppcode}{0.9}
File file("mesh.xml");
Mesh mesh;

file >> mesh;
file << mesh;
\end{cppcode}
Mesh data may be converted to the native \dolfin{} XML format
from Gmsh, Medit, Diffpack, ABAQUS, Exodus~II and
StarCD formats using the conversion utility \emp{dolfin-convert}.

Solution data may be exported in a number of formats, including the
VTK~XML format which is useful for visualizing a \emp{Function} in
VTK-based tools, such as ParaView.
\dolfin{} also provides built-in plotting for \emp{Mesh},
\emp{MeshFunction} and \emp{Function} using Viper \cite{www:Viper} by
\begin{cppcode}{0.9}
plot(mesh);
plot(mesh_function);
plot(u);
\end{cppcode}

\section{Applications}
\label{sec:applications}

We present here a collection of examples to demonstrate the use of
\dolfin{} for solving partial differential equations and related
problems of interest. A more extensive range of examples are
distributed with the \dolfin{} source code. For a particularly
complicated application to reservoir modeling, we refer to
\citeN{wells:2009}. Some issues of particular relevance to solid
mechanics problems, such as plasticity,
are discussed in \citeN{oelgaard:2008c}. All examples correspond to
\dolfin{}~0.9.5.

\subsection{Evaluating functionals}

We begin with the simplest form that we can evaluate, a functional. In
the absence of viscous stresses, the lift acting on a body can be computed
by integrating the pressure multiplied by a suitable component of the
unit vector normal to the surface of interest. The definition of this
functional is shown in Table~\ref{tab:lift_functional}. From this
definition, C++ code may be generated using a form compiler and then
used to compute the lift generated by a computed pressure field.
\begin{table}
\begin{ffccode}{0.9}
element = FiniteElement("Lagrange", triangle, 1)

p = Function(element)
n = triangle.n
M = p*n[1]*ds
\end{ffccode}
\caption{Definition of the functional for computing the lift.}
\label{tab:lift_functional}
\end{table}

%
%
%
%
%
%
%

Another common application of functionals is the evaluation of various
norms or evaluating the error of a computed solution when the exact
solution is known. For example, one may define the squared $L^{2}$ and
$H^{1}_{0}$ norms as \emp{v*v*dx} and \emp{dot(grad(v), grad(v))*dx}
respectively. Alternatively, one may use the built-in \dolfin{}
functions \emp{norm} and \emp{errornorm} to evaluate norms and errors:
\begin{pythoncode}{0.9}
print norm(v, "L2")
print norm(v, "H1")
print norm(v, "H10")
print norm(v, "Hdiv")
...
print errornorm(u_h, u, "L2")
print errornorm(u_h, u, "H1")
print errornorm(u_h, u, "H10")
print errornorm(u_h, u, "Hdiv")
...
\end{pythoncode}

\subsection{Solving linear partial differential equations}

To illustrate the use of \dolfin{} for solving simple linear partial
differential equations, we consider Poisson's equation $-\Delta u =f$
discretized using three different methods: an $H^{1}$-conforming
primal approach using standard continuous Lagrange basis functions; a
mixed method using $H(\Div)\times L^2$-conforming elements; and a
discontinuous Galerkin method using $L^{2}$-conforming
Lagrange basis functions.

\subsubsection{$H^{1}$-conforming discretization of Poisson's equation}

For the standard $H^1$-conforming approach, the bilinear and
linear forms are given by
\begin{align}
  a(v, u) &= \int_{\Omega} \nabla v \cdot \nabla u \dx, \\
  L(v)    &= \int_{\Omega} v f \dx,
\end{align}
and the forms may be specified in \dolfin{} by
\begin{pythoncode}{0.9}
V = FunctionSpace(mesh, "CG", 1)
v = TestFunction(V)
u = TrialFunction(V)
f = Expression(...)

a = dot(grad(v), grad(u))*dx
L = v*f*dx
\end{pythoncode}

\subsubsection{$H(\Div) \times L^2$-conforming discretization of Poisson's equation}

For the mixed version of the Poisson problem, with $u = 0 \ {\rm on} \
\partial\Omega$, the bilinear and linear forms read
\cite{brezzi:book}:
\begin{align}
  a(\tau, w; \sigma, u) &= \int_{\Omega} \tau \cdot \sigma
  - (\nabla \cdot \tau) \, u
  + w \, (\nabla \cdot \sigma) \dx, \\
  L(\tau, w) &= \int_{\Omega} w f \dx,
\end{align}
where $\tau, \sigma \in V$, $w, u \in W$ and
\begin{align}
  V &= \bracc{\tau \in H\brac{{\rm div}, \Omega}:
    \tau\vert_{K} \in [P_q]^n \brac{K} \, \forall K},
  \\
  W &= \bracc{w \in L^{2}\brac{\Omega}:
    w\vert_{K} \in P_{q-1}\brac{K} \, \forall K}.
\end{align}
The corresponding implementation in \dolfin{} for $q = 2$ reads:

\begin{pythoncode}{0.9}
V = FunctionSpace(mesh, "BDM", 2)
W = FunctionSpace(mesh, "DG", 1)

mixed_space = V + W

(tau, w)   = TestFunctions(mixed_space)
(sigma, u) = TrialFunctions(mixed_space)

f = Expression(...)

a = (dot(tau, sigma) - div(tau)*u + w*div(sigma))*dx
L = w*f*dx
\end{pythoncode}

\subsubsection{$L^2$-conforming discretization of Poisson's equation}

For a discontinuous interior penalty formulation of the Poisson problem,
the bilinear and linear forms read:
\begin{multline}
  a(v,u) =
    \int_{\Omega \backslash \Gamma^{0}}\nabla v \cdot \nabla u \dx
  - \int_{\Gamma^{0}} \jump{v} \cdot \avg{\nabla u}  \ds
  - \int_{\Gamma^{0}} \avg{\nabla v} \cdot \jump{u}  \ds
\\
  - \int_{\partial\Omega} v \vect{n} \cdot \nabla u  \ds
  - \int_{\partial\Omega} \nabla v \cdot u \vect{n}  \ds
  + \int_{\Gamma^{0}} \frac{\alpha}{h} \jump{v} \cdot \jump{u}  \ds
  + \int_{\partial\Omega} \frac{\alpha}{h} v u  \ds
\label{eq:poisson_dg_bilinear}
\end{multline}
and
\begin{equation}
  L(v) = \int_{\Omega} v f  \dx,
\label{eq:poisson_dg_linear}
\end{equation}
where $\Gamma^{0}$ denotes all interior facets and $v, u \in V =
\bracc{v \in L^{2}\brac{\Omega}: v|_{K} \in P_q\brac{K} \,
  \forall K}$.  Using {\tt ds} to denote integration over exterior
facets and {\tt dS} to denote integration over interior facets,
the corresponding implementation in \dolfin{}
for $q = 1$ reads as follows:
\begin{pythoncode}{0.9}
V = FunctionSpace(mesh, "DG", 1)

v = TestFunction(V)
u = TrialFunction(V)
f = Expression(...)

n = FacetNormal(mesh)
h = CellSize(mesh)

alpha = 4.0

a =   dot(grad(v), grad(u))*dx \
    - dot(jump(v, n), avg(grad(u)))*dS \
    - dot(avg(grad(v)), jump(u, n))*dS \
    - v*dot(grad(u), n)*ds - dot(grad(v), n)*u*ds \
    + alpha/h('+')*dot(jump(v, n), jump(u, n))*dS \
    + (alpha/h)*v*u*ds

L = v*f*dx
\end{pythoncode}
\subsection{Solving time-dependent partial differential equations}
\label{sec:convection-diffusion}
Unsteady problems can be solved by defining a variational problem to
be solved in each time step. We illustrate this by solving the
convection--diffusion problem
\begin{equation} \label{eq:convdiff}
  \dot{u} + b \cdot \nabla u - \nabla \cdot (c \nabla u) = f.
\end{equation}
The velocity field $b = b(x)$ may be a user-defined expression or an
earlier computed solution. Multiplying~\eqref{eq:convdiff} with a test
function and discretizing in time using the Crank--Nicolson method,
we obtain
\begin{equation} \label{eq:varproblem,convdiff,2}
  \int_{\Omega} v \, (u^{n} - u^{n-1})
  + k_n \, v \, \bar{b} \cdot \nabla \bar{u}
  + k_n \, \bar{c} \nabla v \cdot \nabla \bar{u} \dx
 = \int_{\Omega} k_n \, v \bar{f} \dx,
\end{equation}
where $k_n = t_{n} - t_{n-1}$ is the time step and
$\bar{x} = (x^{n} + x^{n-1})/2$.  We may implement the
problem~\eqref{eq:varproblem,convdiff,2} in \dolfin{} by moving all
terms involving $u^{n-1}$ to the right-hand side. Alternatively, we
may rely on the built-in operators \emp{lhs} and \emp{rhs} to
extract the pair of bilinear and linear forms as illustrated in
Table~\ref{code:convdiff,ffc}. In Table~\ref{code:convdiff,dolfin} we
show the corresponding C++ program.
\begin{table}
\begin{ffccode}{0.9}
scalar = FiniteElement("Lagrange", triangle, 1)
vector = VectorElement("Lagrange", triangle, 2)

v  = TestFunction(scalar)  # test function
u1 = TrialFunction(scalar) # solution at t_n
u0 = Function(scalar)      # solution at t_{n-1}
b  = Function(vector)      # convective velocity
f  = Function(scalar)      # source term
c  = 0.005                 # diffusivity
k  = 0.05                  # time step

u = 0.5*(u0 + u1)
F = v*(u1 - u0)*dx + k*v*dot(b, grad(u))*dx + k*c*dot(grad(v), grad(u))*dx
a = lhs(F)
L = rhs(F) + k*v*f*dx
\end{ffccode}
\caption{Specification of the variational problem for the unsteady
convection-diffusion equation~\eqref{eq:convdiff}.}
\label{code:convdiff,ffc}
\end{table}
\begin{table}
\begin{cppcode}{0.9}
// Read mesh from file
Mesh mesh("mesh.xml.gz");

// Read velocity field from file
Velocity::FunctionSpace W(mesh);
Function velocity(W, "velocity.xml.gz");

// Read sub domain markers from file
MeshFunction<unsigned int> sub_domains(mesh, "subdomains.xml.gz");

// Create function space
ConvectionDiffusion::FunctionSpace V(mesh);

// Create source term and initial condition
Constant f(0);
Function u(V);

// Set up variational forms
ConvectionDiffusion::BilinearForm a(V, V);
a.b = velocity;
ConvectionDiffusion::LinearForm L(V);
L.u0 = u; L.b = velocity; L.f = f;

// Set up boundary condition
Constant g(1);
DirichletBC bc(V, g, sub_domains, 1);

// Linear system
Matrix A;
Vector b;

// Assemble matrix and apply boundary conditions
assemble(A, a);
bc.apply(A);

// Parameters for time-stepping
double T = 2.0; double k = 0.05; double t = k;

// Output file
File file("temperature.pvd");

// Time-stepping
while (t < T)
{
  assemble(b, L);
  bc.apply(b);
  solve(A, u.vector(), b, lu);
  file << u;
  t += k;
}
\end{cppcode}
\caption{Implementation of the solver for the unsteady
convection-diffusion equation~\eqref{eq:convdiff}.}
\label{code:convdiff,dolfin}
\end{table}
\subsection{Solving nonlinear partial differential equations}
\label{sec:nonlinear-pde}
Solution procedures for nonlinear differential equations are
inherently more complex and diverse than those for linear
equations. With this in mind, the design of \dolfin{} allows users to
build complex solution algorithms for nonlinear problems using the
basic building blocks \emp{assemble} and \emp{solve}. However, a
built-in Newton solver is also provided which suffices for many
problems. We illustrate the solution of a nonlinear problem for the
following nonlinear Poisson-like equation:
\begin{align}
  \label{eq:poisson,nonlinear,1}
  - \nabla \cdot \brac{1+ u^{2}} \nabla  u &= f \quad \mbox{ in } \Omega, \\
  \label{eq:poisson,nonlinear,2}
  u &= 0 \quad \mbox{ on } \partial\Omega.
\end{align}
Multiplying by a test function $v \in V = H^1_0(\Omega)$ and
integrating over the domain $\Omega$, we obtain
\begin{equation}
 F(v; u) \equiv \int_{\Omega} \brac{1+ u^{2}} \nabla v \cdot \nabla  u \dx
 - \int_{\Omega} v f = 0,
\label{eqn:nonlinear_functional}
\end{equation}
where we note that~$F : V \times V \rightarrow \R$ is linear in its
first argument and nonlinear in its second argument. To solve the
nonlinear problem by Newton's method, we compute the Gateaux
derivative $D_{u}F(v; u)$ and obtain
\begin{equation}
  \begin{split}
  a(v, \delta u; u) &\equiv D_uF(v; u) \delta u
  = \left. \frac{dF(v; u + \epsilon \delta u)}{d\epsilon} \right|_{\epsilon =0} \\
  &= \int_{\Omega} \brac{1+ u^{2}} \nabla v \cdot \nabla \delta u \dx
  + \int_{\Omega} 2 u \delta u \nabla v \cdot \nabla  u \dx.
  \end{split}
\end{equation}
We note that $F(\cdot; u) : V \rightarrow \R$ is a linear form for
every fixed $u$ and that $a(\cdot, \cdot; u) : V \times V
\rightarrow \R$ is a bilinear form for every fixed $u$. A full
solver
for~\eqref{eq:poisson,nonlinear,1}--\eqref{eq:poisson,nonlinear,2} in
the case $f(x, y) = x \sin y$ is presented in
Table~\ref{code:poisson,nonlinear}.
The form language UFL supports automatic differentiation, so many problems,
including this one, can also be linearized automatically.
\begin{table}
\begin{pythoncode}{0.9}
from dolfin import *

# Create mesh and define function space
mesh = UnitSquare(32, 32)
V = FunctionSpace(mesh, "CG", 1)

# Define boundary condition
bc = DirichletBC(V, Constant(0), DomainBoundary())

# Define source term and solution function
f = Expression("x[0]*sin(x[1])")
u = Function(V)

# Define variational problem
v  = TestFunction(V)
du = TrialFunction(V)
a  = (1.0 + u*u)*dot(grad(v), grad(du))*dx + \
     2*u*du*dot(grad(v), grad(u))*dx
L  = (1.0 + u*u)*dot(grad(v), grad(u))*dx - v*f*dx

# Solve nonlinear variational problem
problem = VariationalProblem(a, L, bc, nonlinear=True)
problem.solve(u)

# Plot solution and solution gradient
plot(u)
plot(grad(u))
\end{pythoncode}
\caption{Implementation of a solver for the nonlinear Poisson
  problem~\eqref{eq:poisson,nonlinear,1}--\eqref{eq:poisson,nonlinear,2}.}
\label{code:poisson,nonlinear}
\end{table}
\section{Conclusions}
\label{sec:conclusions}
We have presented a problem solving environment that largely automates
the finite element approximation of solutions to differential
equations. This is achieved by generating computer code for parts of
the problem which are specific to the considered differential
equation, and designing a generic library which reflects the
mathematical structure of finite element variational problems. Using a
high level of mathematical abstraction and automated code generation,
the system can be designed for both readability and performance,
allowing new models to be implemented rapidly and solved efficiently.

Until recently, the focus has been on automating the assembly of
linear systems arising from the finite element discretization of
variational problems, in particular with regards to providing a
general implementation independent of the variational problem, the
mesh, the discretizing finite element space(s) and the linear algebra
backend.  More recently,  efficient
parallel computing has been  added and
automated error estimation/adaptivity is being developed.
\begin{ack}
We acknowledge the contributions that many people have made to the
development of \dolfin{}. Johan Hoffman and Johan Jansson both
contributed to early versions of \dolfin{}, in particular with
algorithms for adaptive mesh refinement and solution of ordinary
differential equations. Martin Aln{\ae}s, Kent-Andre Mardal and Ola
Skavhaug have been involved in the design and implementation of the
\dolfin{} linear algebra interfaces and backends. Johan Hake and
Ola Skavhaug have made significant contributions to the design of the
\dolfin{} Python interface. Johannes Ring and Ilmar Wilbers maintain
the \dolfin{} build system and produce packages for various
platforms. We also mention Benjamin Kehlet, Gustav Magnus Vikstr\o{}m,
Kristian \O{}lgaard, Niclas Jansson, Dag Lindbo, \AA{}smund
\O{}degard, Evan Lezar and Shawn Walker.\footnote{Many more
  people have contributed patches. We list here only those who have
  contributed more than ca 10~patches but acknowledge the importance of
  all contributions.}

AL is supported by an Outstanding Young Investigator grant from the
Research Council of Norway, NFR 180450.
\end{ack}
\bibliographystyle{acmtrans}
\bibliography{references}

\begin{thebibliography}{}

\bibitem[\protect\citeauthoryear{Aln{\ae}s}{Aln{\ae}s}{2009}]{Alnaes2009}
{\sc Aln{\ae}s, M.~S.} 2009.
\newblock A compiler framework for automatic linearization and efficient
  discretization of nonlinear partial differential equations.
\newblock Ph.D. thesis, University of Oslo.
\newblock
  \url{http://simula.no/research/sc/publications/Simula.SC.626/simula_pdf_file%
}.

\bibitem[\protect\citeauthoryear{Aln\ae{}s, Langtangen, Logg, Mardal, and
  Skavhaug}{Aln\ae{}s et~al\mbox{.}}{2009}]{www:UFC}
{\sc Aln\ae{}s, M.~S.}, {\sc Langtangen, H.~P.}, {\sc Logg, A.}, {\sc Mardal,
  K.-A.}, {\sc and} {\sc Skavhaug, O.} 2009.
\newblock {UFC}.
\newblock \url{http://www.fenics.org/wiki/UFC/}.

\bibitem[\protect\citeauthoryear{Aln\ae{}s and Logg}{Aln\ae{}s and
  Logg}{2009}]{www:UFL}
{\sc Aln\ae{}s, M.~S.} {\sc and} {\sc Logg, A.} 2009.
\newblock {UFL}.
\newblock \url{http://www.fenics.org/wiki/UFL/}.

\bibitem[\protect\citeauthoryear{Aln{\ae}s and Mardal}{Aln{\ae}s and
  Mardal}{2009}]{www:SyFi}
{\sc Aln{\ae}s, M.~S.} {\sc and} {\sc Mardal, K.-A.} 2009.
\newblock {\em {SyFi}}.
\newblock \url{http://www.fenics.org/wiki/SyFi/}.

\bibitem[\protect\citeauthoryear{Aln{\ae}s and Mardal}{Aln{\ae}s and
  Mardal}{2010}]{alnaes:2009}
{\sc Aln{\ae}s, M.~S.} {\sc and} {\sc Mardal, K.-A.} 2010.
\newblock On the efficiency of symbolic computations combined with code
  generation for finite element methods.
\newblock {\em ACM Transactions on Mathematical Software\/}~{\em 37},
  6:1--6:26.

\bibitem[\protect\citeauthoryear{Aln\ae{}s, Mardal, and Westlie}{Aln\ae{}s
  et~al\mbox{.}}{2009}]{www:Instant}
{\sc Aln\ae{}s, M.~S.}, {\sc Mardal, K.-A.}, {\sc and} {\sc Westlie, M.} 2009.
\newblock Instant.
\newblock \url{http://www.fenics.org/wiki/Instant}.

\bibitem[\protect\citeauthoryear{Balay, Buschelman, Gropp, Kaushik, Knepley,
  McInnes, Smith, and Zhang}{Balay et~al\mbox{.}}{2009}]{petsc:www}
{\sc Balay, S.}, {\sc Buschelman, K.}, {\sc Gropp, W.~D.}, {\sc Kaushik, D.},
  {\sc Knepley, M.~G.}, {\sc McInnes, L.~C.}, {\sc Smith, B.~F.}, {\sc and}
  {\sc Zhang, H.} 2009.
\newblock {PETSc} {W}eb page.
\newblock \url{http://www.mcs.anl.gov/petsc/}.

\bibitem[\protect\citeauthoryear{Bangerth, Hartmann, and Kanschat}{Bangerth
  et~al\mbox{.}}{2007}]{bangerth:2007}
{\sc Bangerth, W.}, {\sc Hartmann, R.}, {\sc and} {\sc Kanschat, G.} 2007.
\newblock deal.{II} --- {A} general purpose object oriented finite element
  library.
\newblock {\em {ACM} Transactions on Mathematical Software\/}~{\em 33,\/}~4,
  24.

\bibitem[\protect\citeauthoryear{Beazley}{Beazley}{2003}]{beazley:2003}
{\sc Beazley, D.~M.} 2003.
\newblock Automated scientific software scripting with {SWIG}.
\newblock {\em Future Generation Computer Systems\/}~{\em 19,\/}~5, 599--609.

\bibitem[\protect\citeauthoryear{Berti}{Berti}{2002}]{Ber02}
{\sc Berti, G.} 2002.
\newblock Generic programming for mesh algorithms: {T}owards universally usable
  geometric components.
\newblock In {\em Proceedings of the Fifth World Congress on Computational
  Mechanics (WCCM~V)}, {H.~A. Mang}, {F.~G. Rammerstorfer}, {and}
  {J.~Eberhardsteiner}, Eds. Vienna University of Technology, Vienna.
\newblock \url{http://wccm.tuwien.ac.at/publications/Papers/fp81327.pdf}.

\bibitem[\protect\citeauthoryear{Berti}{Berti}{2006}]{Ber06}
{\sc Berti, G.} 2006.
\newblock Gr{AL} -- {T}he grid algorithms library.
\newblock {\em Future Generation Computer Systems\/}~{\em 22}.

\bibitem[\protect\citeauthoryear{Brezzi, Douglas, Fortin, and Marini}{Brezzi
  et~al\mbox{.}}{1987}]{BreDou87}
{\sc Brezzi, F.}, {\sc Douglas, Jr., J.}, {\sc Fortin, M.}, {\sc and} {\sc
  Marini, L.~D.} 1987.
\newblock Efficient rectangular mixed finite elements in two and three space
  variables.
\newblock {\em RAIRO -- Analyse Numerique -- Numerical Analysis\/}~{\em
  21,\/}~4, 581--604.

\bibitem[\protect\citeauthoryear{Brezzi, Douglas, and Marini}{Brezzi
  et~al\mbox{.}}{1985}]{BreDou85}
{\sc Brezzi, F.}, {\sc Douglas, Jr., J.}, {\sc and} {\sc Marini, L.~D.} 1985.
\newblock Two families of mixed finite elements for second order elliptic
  problems.
\newblock {\em Numerische Mathematik\/}~{\em 47,\/}~2, 217--235.

\bibitem[\protect\citeauthoryear{Brezzi and Fortin}{Brezzi and
  Fortin}{1991}]{brezzi:book}
{\sc Brezzi, F.} {\sc and} {\sc Fortin, M.} 1991.
\newblock {\em Mixed and Hybrid Finite Element Methods}. Springer Series in
  Computational Mathematics, vol.~15.
\newblock Springer, New York.

\bibitem[\protect\citeauthoryear{Chen, Davis, Hager, and Rajamanickam}{Chen
  et~al\mbox{.}}{2008}]{chen:2008}
{\sc Chen, Y.}, {\sc Davis, T.~A.}, {\sc Hager, W.~W.}, {\sc and} {\sc
  Rajamanickam, S.} 2008.
\newblock Algorithm 887: {CHOLMOD}, supernodal sparse {C}holesky factorization
  and update/downdate.
\newblock {\em ACM Transactions on Mathematical Software\/}~{\em 35,\/}~3,
  1--14.

\bibitem[\protect\citeauthoryear{Crouzeix and Raviart}{Crouzeix and
  Raviart}{1973}]{CroRav73}
{\sc Crouzeix, M.} {\sc and} {\sc Raviart, P.~A.} 1973.
\newblock Conforming and nonconforming finite element methods for solving the
  stationary stokes equations.
\newblock {\em RAIRO -- Analyse Numerique -- Numerical Analysis\/}~{\em 7},
  33--76.

\bibitem[\protect\citeauthoryear{Davis}{Davis}{2004}]{davis:2004}
{\sc Davis, T.~A.} 2004.
\newblock Algorithm 832: {UMFPACK} v4.3---an unsymmetric-pattern multifrontal
  method.
\newblock {\em ACM Transactions on Mathematical Software\/}~{\em 30,\/}~2,
  196--199.

\bibitem[\protect\citeauthoryear{Dular, Geuzaine, et~al\mbox{.}}{Dular
  et~al\mbox{.}}{2009}]{getdp:www}
{\sc Dular, P.}, {\sc Geuzaine, C.}, {\sc et~al\mbox{.}} 2009.
\newblock {GetDP}: {A} general environment for the treatment of discrete
  problems.
\newblock \url{http://geuz.org/getdp/}.

\bibitem[\protect\citeauthoryear{FEniCS}{FEniCS}{2009}]{fenics:www}
{\sc FEniCS}. 2009.
\newblock {FE}ni{CS} {P}roject.
\newblock \url{http://www.fenics.org/}.

\bibitem[\protect\citeauthoryear{Gottschling and Lumsdaine}{Gottschling and
  Lumsdaine}{2009}]{mtl4:www}
{\sc Gottschling, P.} {\sc and} {\sc Lumsdaine, A.} 2009.
\newblock The {M}atrix {T}emplate {L}ibrary 4.
\newblock \url{http://www.osl.iu.edu/research/mtl/mtl4/}.

\bibitem[\protect\citeauthoryear{??}{GTS}{2009}]{gts:www}
GTS 2009.
\newblock {GNU} {T}riangulated {S}urface {L}ibrary ({GTS}).
\newblock \url{http://gts.sourceforge.net/}.

\bibitem[\protect\citeauthoryear{Heroux, Bartlett, Howle, Hoekstra, Hu, Kolda,
  Lehoucq, Long, Pawlowski, Phipps, Salinger, Thornquist, Tuminaro,
  Willenbring, Williams, and Stanley}{Heroux
  et~al\mbox{.}}{2005}]{trilinos:2005}
{\sc Heroux, M.~A.}, {\sc Bartlett, R.~A.}, {\sc Howle, V.~E.}, {\sc Hoekstra,
  R.~J.}, {\sc Hu, J.~J.}, {\sc Kolda, T.~G.}, {\sc Lehoucq, R.~B.}, {\sc Long,
  K.~R.}, {\sc Pawlowski, R.~P.}, {\sc Phipps, E.~T.}, {\sc Salinger, A.~G.},
  {\sc Thornquist, H.~K.}, {\sc Tuminaro, R.~S.}, {\sc Willenbring, J.~M.},
  {\sc Williams, A.}, {\sc and} {\sc Stanley, K.~S.} 2005.
\newblock An overview of the {T}rilinos project.
\newblock {\em {ACM} Transactions on Mathematical Software\/}~{\em 31,\/}~3,
  397--423.

\bibitem[\protect\citeauthoryear{Kirby}{Kirby}{2004}]{Kir04}
{\sc Kirby, R.~C.} 2004.
\newblock Algorithm 839: {FIAT}, a new paradigm for computing finite element
  basis functions.
\newblock {\em {ACM} Transactions on Mathematical Software\/}~{\em 30,\/}~4,
  502--516.

\bibitem[\protect\citeauthoryear{Kirby}{Kirby}{2009}]{www:FIAT}
{\sc Kirby, R.~C.} 2009.
\newblock {\em {FIAT}}.
\newblock \url{http://www.fenics.org/fiat/}.

\bibitem[\protect\citeauthoryear{Kirby, Knepley, Logg, and Scott}{Kirby
  et~al\mbox{.}}{2005}]{kirby:2005}
{\sc Kirby, R.~C.}, {\sc Knepley, M.~G.}, {\sc Logg, A.}, {\sc and} {\sc Scott,
  L.~R.} 2005.
\newblock Optimizing the evaluation of finite element matrices.
\newblock {\em SIAM Journal on Scientific Computing\/}~{\em 27,\/}~3, 741--758.

\bibitem[\protect\citeauthoryear{Kirby and Logg}{Kirby and
  Logg}{2006}]{kirby:2006}
{\sc Kirby, R.~C.} {\sc and} {\sc Logg, A.} 2006.
\newblock A compiler for variational forms.
\newblock {\em {ACM} Transactions on Mathematical Software\/}~{\em 32,\/}~3,
  417--444.

\bibitem[\protect\citeauthoryear{Kirby and Logg}{Kirby and
  Logg}{2007}]{logg:article:11}
{\sc Kirby, R.~C.} {\sc and} {\sc Logg, A.} 2007.
\newblock Efficient compilation of a class of variational forms.
\newblock {\em {ACM} Transactions on Mathematical Software\/}~{\em 33,\/}~3.

\bibitem[\protect\citeauthoryear{Kirby, Logg, Scott, and Terrel}{Kirby
  et~al\mbox{.}}{2006}]{kirby:2006b}
{\sc Kirby, R.~C.}, {\sc Logg, A.}, {\sc Scott, L.~R.}, {\sc and} {\sc Terrel,
  A.~R.} 2006.
\newblock Topological optimization of the evaluation of finite element
  matrices.
\newblock {\em SIAM Journal on Scientific Computing\/}~{\em 28,\/}~1, 224--240.

\bibitem[\protect\citeauthoryear{Knepley and Karpeev}{Knepley and
  Karpeev}{2009}]{KnepleyKarpeev07A}
{\sc Knepley, M.~G.} {\sc and} {\sc Karpeev, D.~A.} 2009.
\newblock Mesh algorithms for {PDE} with {S}ieve {I}: Mesh distribution.
\newblock {\em Scientific Programming\/}~{\em 17,\/}~3, 215--230.

\bibitem[\protect\citeauthoryear{Langtangen}{Langtangen}{2003}]{langtangen:boo%
k}
{\sc Langtangen, H.~P.} 2003.
\newblock {\em Computational Partial Differential Equations: Numerical Methods
  and Diffpack Programming}. Texts in Computational Science and Engineering,
  vol.~1.
\newblock Springer.

\bibitem[\protect\citeauthoryear{Logg}{Logg}{2007}]{Logg2007a}
{\sc Logg, A.} 2007.
\newblock Automating the finite element method.
\newblock {\em Arch. Comput. Methods Eng.\/}~{\em 14,\/}~2, 93--138.

\bibitem[\protect\citeauthoryear{Logg}{Logg}{2009}]{logg:2008}
{\sc Logg, A.} 2009.
\newblock Efficient representation of computational meshes.
\newblock {\em International Journal of Computational Science and
  Engineering\/}~{\em 4,\/}~4, 283--295.

\bibitem[\protect\citeauthoryear{Logg, {\O}lgaard, Rognes, Wells,
  et~al\mbox{.}}{Logg et~al\mbox{.}}{2009}]{www:ffc}
{\sc Logg, A.}, {\sc {\O}lgaard, K.~B.}, {\sc Rognes, M.~E.}, {\sc Wells,
  G.~N.}, {\sc et~al\mbox{.}} 2009.
\newblock {\em {FFC}}.
\newblock \url{http://www.fenics.org/ffc/}.

\bibitem[\protect\citeauthoryear{Long et~al\mbox{.}}{Long
  et~al\mbox{.}}{2009}]{sundance:www}
{\sc Long, K.} {\sc et~al\mbox{.}} 2009.
\newblock Sundance.
\newblock \url{http://www.math.ttu.edu/~klong/Sundance/html/}.

\bibitem[\protect\citeauthoryear{N{\'e}d{\'e}lec}{N{\'e}d{\'e}lec}{1980}]{Ned8%
0}
{\sc N{\'e}d{\'e}lec, J.-C.} 1980.
\newblock Mixed finite elements in {${\bf R}\sp{3}$}.
\newblock {\em Numerische Mathematik\/}~{\em 35,\/}~3, 315--341.

\bibitem[\protect\citeauthoryear{Nikbakht and Wells}{Nikbakht and
  Wells}{2009}]{nikbakht:2009}
{\sc Nikbakht, M.} {\sc and} {\sc Wells, G.~N.} 2009.
\newblock Automated modelling of evolving discontinuities.
\newblock {\em Algorithms\/}~{\em 2,\/}~3, 1008--1030.

\bibitem[\protect\citeauthoryear{{\O}lgaard, Logg, and Wells}{{\O}lgaard
  et~al\mbox{.}}{2008}]{oelgaard:2008}
{\sc {\O}lgaard, K.~B.}, {\sc Logg, A.}, {\sc and} {\sc Wells, G.~N.} 2008.
\newblock Automated code generation for discontinuous {G}alerkin methods.
\newblock {\em SIAM Journal on Scientific Computing\/}~{\em 31,\/}~2, 849--864.

\bibitem[\protect\citeauthoryear{{\O}lgaard and Wells}{{\O}lgaard and
  Wells}{2010}]{oelgaard:2009}
{\sc {\O}lgaard, K.~B.} {\sc and} {\sc Wells, G.~N.} 2010.
\newblock Optimisations for quadrature representations of finite element
  tensors through automated code generation.
\newblock {\em ACM Transactions on Mathematical Software\/}~{\em 37,\/}~1,
  8:1--8:23.

\bibitem[\protect\citeauthoryear{{\O}lgaard, Wells, and Logg}{{\O}lgaard
  et~al\mbox{.}}{2008}]{oelgaard:2008c}
{\sc {\O}lgaard, K.~B.}, {\sc Wells, G.~N.}, {\sc and} {\sc Logg, A.} 2008.
\newblock Automated computational modelling for solid mechanics.
\newblock In {\em IUTAM Symposium on Theoretical, Computational and Modelling
  Aspects of Inelastic Media}, {B.~D. Reddy}, Ed. IUTAM Bookseries, vol.~11.
  Springer, 195--204.

\bibitem[\protect\citeauthoryear{Pironneau, Hecht, and Le~Hyaric}{Pironneau
  et~al\mbox{.}}{2009}]{freefem:www}
{\sc Pironneau, O.}, {\sc Hecht, F.}, {\sc and} {\sc Le~Hyaric, A.} 2009.
\newblock {FreeFEM++}.
\newblock \url{http://www.freefem.org/}.

\bibitem[\protect\citeauthoryear{Prud'homme}{Prud'homme}{2007}]{prudhomme2008}
{\sc Prud'homme, C.} 2007.
\newblock Life: {O}verview of a unified {C++} implementation of the finite and
  spectral element methods in {1D}, {2D} and {3D}.
\newblock In {\em Applied Parallel Computing. State of the Art in Scientific
  Computing}. Lecture Notes in Computer Science, vol. 4699/2009. Springer
  Berlin / Heidelberg, 712--721.

\bibitem[\protect\citeauthoryear{Prud'homme}{Prud'homme}{2009}]{life:www}
{\sc Prud'homme, C.} 2009.
\newblock Life.
\newblock \url{http://ljkforge.imag.fr/life}.

\bibitem[\protect\citeauthoryear{Raviart and Thomas}{Raviart and
  Thomas}{1977}]{RavTho77b}
{\sc Raviart, P.-A.} {\sc and} {\sc Thomas, J.~M.} 1977.
\newblock Primal hybrid finite element methods for {$2$}nd order elliptic
  equations.
\newblock {\em Mathematics of Computation\/}~{\em 31,\/}~138, 391--413.

\bibitem[\protect\citeauthoryear{Rivara}{Rivara}{1991}]{rivara:1991}
{\sc Rivara, M.-C.} 1991.
\newblock Local modification of meshes for adaptive and/or multigrid
  finite-element methods.
\newblock {\em Journal of Computational and Applied Mathematics\/}~{\em
  36,\/}~1, 79 -- 89.

\bibitem[\protect\citeauthoryear{Rognes, Kirby, and Logg}{Rognes
  et~al\mbox{.}}{2009}]{rognes:2008}
{\sc Rognes, M.~E.}, {\sc Kirby, R.~C.}, {\sc and} {\sc Logg, A.} 2009.
\newblock Efficient assembly of {$H({\rm div})$} and {$H({\rm curl})$}
  conforming finite elements.
\newblock {\em SIAM Journal on Scientific Computing\/}~{\em 31,\/}~6,
  4130--4151.

\bibitem[\protect\citeauthoryear{Skavhaug}{Skavhaug}{2009}]{www:Viper}
{\sc Skavhaug, O.} 2009.
\newblock Viper.
\newblock \url{http://www.fenics.org/wiki/Viper}.

\bibitem[\protect\citeauthoryear{??}{SWIG}{2009}]{swig:www}
SWIG 2009.
\newblock {S}implified {W}rapper and {I}nterface {G}enerator {(SWIG)}.
\newblock \url{http://www.swig.org/}.

\bibitem[\protect\citeauthoryear{Walter, Koch, et~al\mbox{.}}{Walter
  et~al\mbox{.}}{2009}]{ublas:www}
{\sc Walter, J.}, {\sc Koch, M.}, {\sc et~al\mbox{.}} 2009.
\newblock {uBLAS}.
\newblock \url{http://www.boost.org/}.

\bibitem[\protect\citeauthoryear{Wells, Hooijkaas, and Shan}{Wells
  et~al\mbox{.}}{2008}]{wells:2009}
{\sc Wells, G.~N.}, {\sc Hooijkaas, T.}, {\sc and} {\sc Shan, X.} 2008.
\newblock Modelling temperature effects on multiphase flow through porous
  media.
\newblock {\em Philosophical Magazine\/}~{\em 88,\/}~28--29, 3265--3279.

\end{thebibliography}
\begin{received}
\end{received}
\end{document}